\documentclass[5pt,letterpaper]{article}

\usepackage{color}
\usepackage{upgreek}
\usepackage{amsmath}
\usepackage{amssymb}
\usepackage{graphicx}
\usepackage{epstopdf}

\usepackage{cite}

\title{\textbf{Plasmon-polaritonic quadrupole topological insulators}}

\author{Ying Chen$^1$, Zhi-Kang Lin$^2$, Huanyang Chen$^{1,\dag}$, Jian-Hua Jiang$^{2,\dag}$}
\date{}
\begin{document}
\maketitle
$^1$Institute of Electromagnetics and Acoustics and Key Laboratory of Electromagnetic Wave Science and Detection Technology, Xiamen University, Xiamen 361005, China.

$^2$School of Physical Science and Technology, and Collaborative Innovation Center of Suzhou Nano Science and Technology, Soochow University, 1 Shizi Street, Suzhou, 215006, China.

\vspace{2ex}  

\noindent{$^\dag$Correspondence and requests for materials should be addressed to} $\\$ 
$\color{blue}\textup{kenyon@xmu.edu.cn(Huanyang Chen) and jianhuajiang@suda.edu.cn(Jian-Hua Jiang)}$


\vspace{3ex}  
\begin{abstract}
Quadrupole topological insulator is a symmetry-protected higher-order topological phase with intriguing topology of Wannier bands, which, however, has not yet been realized in plasmonic metamaterials. Here, we propose a lattice of plasmon-polaritonic nanocavities which can realize quadrupole topological insulators by exploiting the geometry-dependent sign-reversal of the couplings between the daisy-like nanocavities. The designed system exhibits various topological and trivial phases as characterized by the nested Wannier bands and the topological quadrupole moment which can be controlled by the distances between the nanocavities. Our study opens a pathway toward plasmonic topological metamaterials with quadrupole topology.
\end{abstract}
%%%%%%%%%%%%%%%%%%%%%%% References %%%%%%%%%%%%%%%%%%%%%%%%%
%\begin{thebibliography}{99}%
%\bibitem{2002PRBFan} S. H. Fan and J. D. Joannopoulos, ``Analysis of guided resonances in photonic crystal slabs,'' Physical Review B. {\bf 65}(23), 235112 (2002).%

%\end{thebibliography}%

%%%%%%%%%%%%%%%%%%%%%%%%%%  body  %%%%%%%%%%%%%%%%%%%%%%%%%%
\section{Introduction.}
Higher-order topological insulators (HOTIs) [1-26] provide a new platform for the study of topological photonics, an emergent branch of science in search of topological phenomena and their applications in photonics [27-45]. HOTIs support multidimensional topological boundary states in a single physical system [11, 18, 26], e.g., the edge and corner states in 2D photonic HOTIs, which can be exploited as topological waveguides and cavity modes in integrated photonic chips. A prototype of HOTIs is the quadrupole topological insulator (QTI) which exhibits gapped edge states and in-gap corner states [1,2, 7-9]. The QTIs are fundamentally different from conventional topological insulators [27, 28] due to its quantized quadrupole moment and the topology of Wannier bands. The Benalcazar-Bernevig-Hughes (BBH) model is the first theoretical model [1, 2] for the QTIs, which is recently realized in mechanical metamaterials [7], electric circuits [8, 9], and coupled ring resonators [17].

On the other hand, plasmon-polaritonic systems offer versatile and effective ways to guide light and to control various photonic modes and their couplings. Plasmonics provides important applications in a broad range of disciplines in photonics with the concepts of plasmonic metamaterials and plasmonic nano-photonics [46-48]. Recently, plasmonic metamaterials have been shown to host conventional topological insulator phases of photons [49-52]. However, higher-order topological phases of photons in plasmonic systems are yet to be studied.

In this Letter, we propose a scheme toward plasmon-polaritonic QTIs. By utilizing the geometry-dependent sign-reversal of the couplings between daisy-like plasmonic modes, we are able to realize the BBH model for the QTIs in plasmonic systems. Various topological and trivial phases can be realized by tuning the distances between the daisy-like cavities, exhibiting versatile topological phenomena related to Wannier bands and topological edge polarizations. Our study opens a pathway toward higher-order quadrupole topological phenomena in plasmonic metamaterials.

\section{Results.}

The designed photonic systems consist of a square lattice of daisy-like cavities embedded into a metallic background as illustrated in Fig. 1(a). The unit-cell consists of two six-petaled cavities filled with air and two eight-petaled cavities filled with a dielectric material of permittivity $\varepsilon_2$=1.525. The lattice constant $a$ is fixed as 1050 nm in our model. The structure of the unit-cell is shown in Fig.1 (b). The daisy-like geometry can be described analytically in polar coordinates. Explicitly, $r_1(\theta)=r_{01}+d_1$cos$(6\theta)$for the six-petaled cavity, while $r_2(\theta)=r_{02}+d_2$cos$(8\theta)$ for the eight-petaled cavity. We adopt $r_{01} = 0.17a$, $d_{1} = 0.062a$, $r_{02} = 0.155a$, and $d_{2} = 0.095a$.
The dashed circles in Fig. 1(a) thus have the inner and outer radii of $r_{01}$ - $d_1$ and $r_{01}$ + $d_1$ ( $r_{02}$ - $d_2$ and $r_{02}$ + $d_2$) for the six-petaled (eight-petaled) cavity, respectively. The above material and geometry parameters are designed to ensure the frequency resonance between the hexagonal mode in the six-petaled cavity and the octopolar mode in the eight-petaled cavity (see Appendix A for details). The electric field profiles for these modes are shown in Fig. 1(a). The permittivity of the metallic background is modeled by a Drude-like form to induce the plasmon polaritons, $\varepsilon_b = 1-{\omega_\textup{p}^2}/{\omega^2}$ where $\omega_\textup{p}$ = 1.37$\times10^{16} s^{-1}$ (adopted from silver’s plasmon frequency [Johnson and Christy, 1972]) [53]. The Drude loss, which induces the non-Hermitian effect, is ignored here, since it has been partly studied elsewhere [54] and is beyond the scope of this work.

The center of this work is to show that the BBH model of QTIs [illustrated in Fig. 1(b)]  can be realized in plasmon-polaritonic lattice systems by exploiting the geometry-dependent sign-reversal of couplings between the daisy-like cavities. Such tunable tight-binding couplings between the cavities rely essentially on the evanescent-wave nature of the plamon polaritons. To realize the BBH model, it is important to achieve the tight-binding configuration in Fig. 1(c). The main challenge is to realize simultaneously both the negative and positive couplings.

We now show that the sign of the coupling between the daisy-like cavities can be well-controlled by their orientations. Computationally, such a coupling can be determined by half of the difference between the frequencies of the even $(\omega_e)$ and the odd $(\omega_o)$ hybridized modes of the two coupled cavities, i.e., $t = \frac{1}{2}(\omega_e-\omega_o)$. For instance, as shown in Fig. 1(d), the coupling along the $x$-direction between the six-petaled cavities is negative, since the the odd mode has higher frequency than the even mode. It turns out that the coupling along the $y$-direction between the six-petaled cavities has opposite sign. Such sign-reversal of the coupling is an intriguing phenomenon discovered in Ref. [55]. In fact, the sign of the coupling depends on whether it is a pole-pole orientation (negative coupling) or node-node orientation (positive coupling) for the two coupled cavities [see Fig. 1(d)]. Exploiting such a mechanism, we can realize exactly the BBH model for QTIs.

We find from first-principle calculations that the coupling strengths between the nearest-neighbor cavities have exponential dependences on the center-to-center distance between them [see Appendex B for details]. Such exponential dependences are signatures of evanescent wave couplings which justify the tight-binding approximation. On the other hand, such dependences enable control of the relative strength between the intra-unit-cell couplings ($\gamma_x$ and $\gamma_y$) and the inter-unit-cell couplings ($\lambda_x$ and $\lambda_y$) through the intra-unit-cell distances ($d_\textup{intra}^x$ and $d_\textup{intra}^y$) and the inter-unit-cell distances ($d_\textup{inter}^x = a - d_\textup{intra}^x$) and ($d_\textup{inter}^y = a - d_\textup{intra}^y$). Fig. 2(a) presents the evolution of the tight-binding coefficients with the ratio $ d_\textup{intra}/d_\textup{inter}$ when $ d_\textup{intra}^x =  d_\textup{intra}^y \equiv {d_\textup{intra}}$ with $d_\textup{inter} = a- d_\textup{intra}$. The figure shows that the tight-binding coefficients can be effectively controlled by the geometry parameters. In particular, the sign-reversal of the couplings $\gamma'_x = -\gamma_x$ and $\lambda'_x = -\lambda_x$ holds for all geometry parameters. With such geometric control, the relative strength of the intra-unit-cell coupling and the inter-unit-cell coupling can be switched. We thus determine the topological phase diagram of the plasmon-polaritonic lattice using the phase diagram of the BBH model [2, 12].

\begin{figure}[h]
	\centering
	\includegraphics[height=5.5cm]{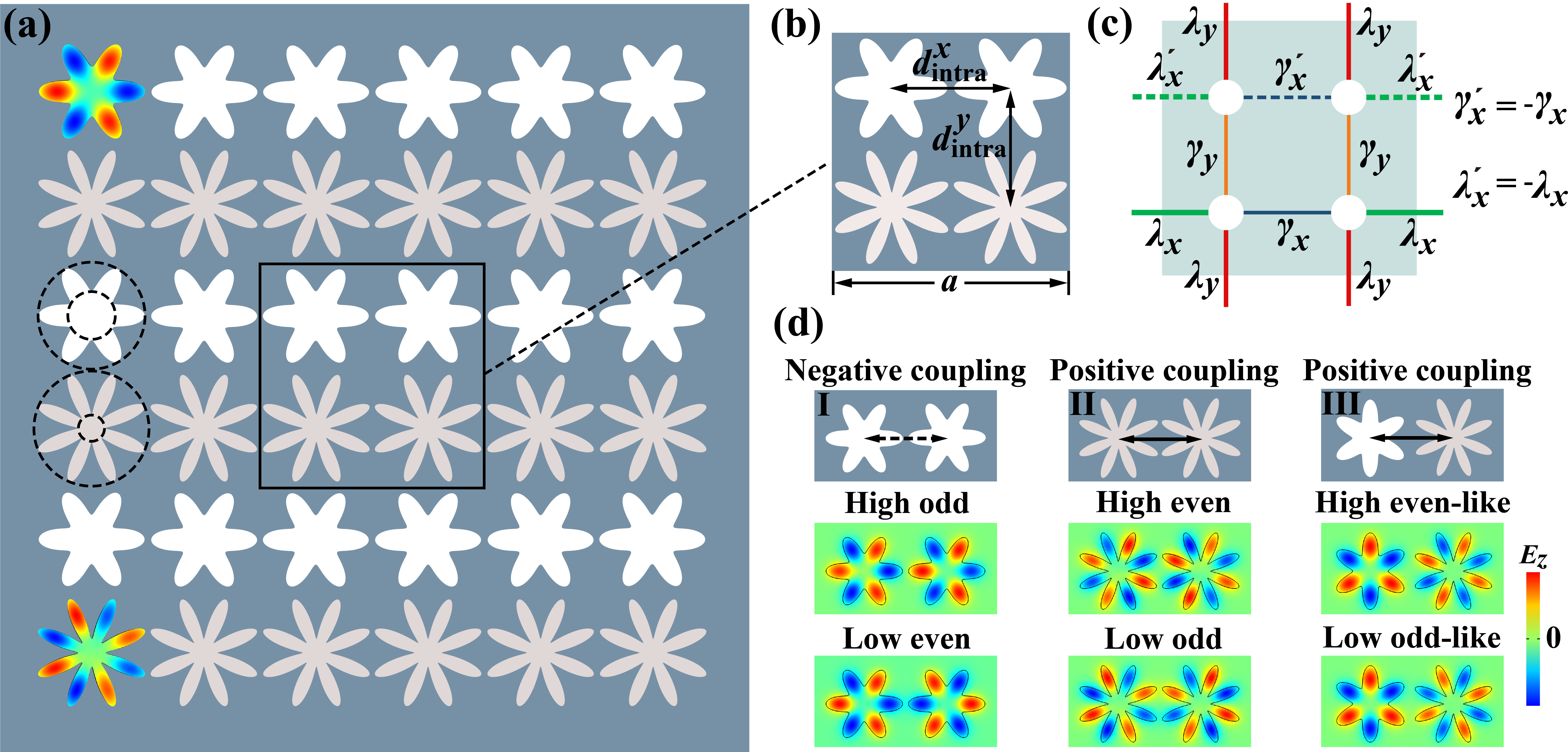}
	\caption{(a) Structure of the designed plasmon-polaritonic system that realizes the BBH tight-binding model for the QTI [illustrated in right panel (c)]. The square-lattice system consists of two types of daisy-like cavities: the six-petaled and the eight-petaled cavities. Left-up (left-down) corner: the electric field profile for the hexagonal (octupole) resonance in the six-petaled (eight-petaled) cavities. (b) The unit-cell structure with lattice constant $a = 1050$ nm. The tunable geometry parameters are the intra-unit-cell distances, $d_\textup{intra}^x$ and $d_\textup{intra}^y$, along the $x$ and $y$ directions, respectively. (c) The BBH tight-binding model for the QTI. The unit-cell is illustrated by the gray region. (d) Illustration of the pole-pole orientation (negative coupling) or node-node orientations (positive coupling).}
	\label{fg1}
\end{figure}

As shown in Fig. 2(b), the topological transitions in our plasmon-polaritonic system take place at the two lines $d_\textup{intra}^x = d_\textup{inter}^x$ and $d_\textup{intra}^y = d_\textup{inter}^y$. Four different phases are separated by the two lines as shown in Fig. 2(b). Those phases are characterized by their topological edge polarizations $\vec{p}^E = (p_x^{E_y}, p_y^{E_x})$ where $p_y^{E_x}$ ($p_x^{E_y}$) represents the polarization along the $y (x)$ direction for the edges normal to the $x (y)$ direction. The QTI has $\vec{p}^E = (\frac{1}{2},\frac{1}{2})$, while the trivial phase without edge or corner state has $\vec{p}^E = (0,0)$. There are other two first-order weak topological insulators with  $\vec{p}^E = (\frac{1}{2},0)$ and $\vec{p}^E = (0,\frac{1}{2})$,  respectively. They have edge states only along the x or the y direction, respectively. The quadrupole topological index is related to the bulk-induced edge polarization as $q_{xy}=2p_x^{E_y} p_y^{E_x}$. Therefore, $q_{xy}$ is nontrivial only in the $\vec{p}^E = (\frac{1}{2},\frac{1}{2})$ phase.  

To characterize the quadrupole topology from the Berry phases, we calculate the Wannier bands and the nested Wannier bands [1, 2]. The Wannier bands are defined through the Wilson-loop operators as follows. The Wilson-loop operator along the y direction is $\hat{W}_{y,\bf{k}}(k_x) = \zeta_{P}   \textup{exp}[i\oint{\hat{A}^y({\bf{k}})dk_y}]$ where the subscripts $y$ and $\bf{k}$ denote the direction and the starting point of the loop, respectively. $\hat{A}^y({\bf{k}})$ is the Berry connection matrix where $A_{nm}^y({\bf{k}}) = i\langle{u_m}({\bf{k}})|\partial_{k_y}|u_n({\bf{k}})\rangle$ with $|u_n(\bf{k})\rangle$ being the periodic part of the Bloch wavefunction for the $n$-th band with the wavevector $\bf{k}$. Here, $\zeta_P$ represents the path-ordering operator along a closed loop in the Brillouin zone (BZ).
\begin{figure}[h]
	\centering
	\includegraphics[height=10cm]{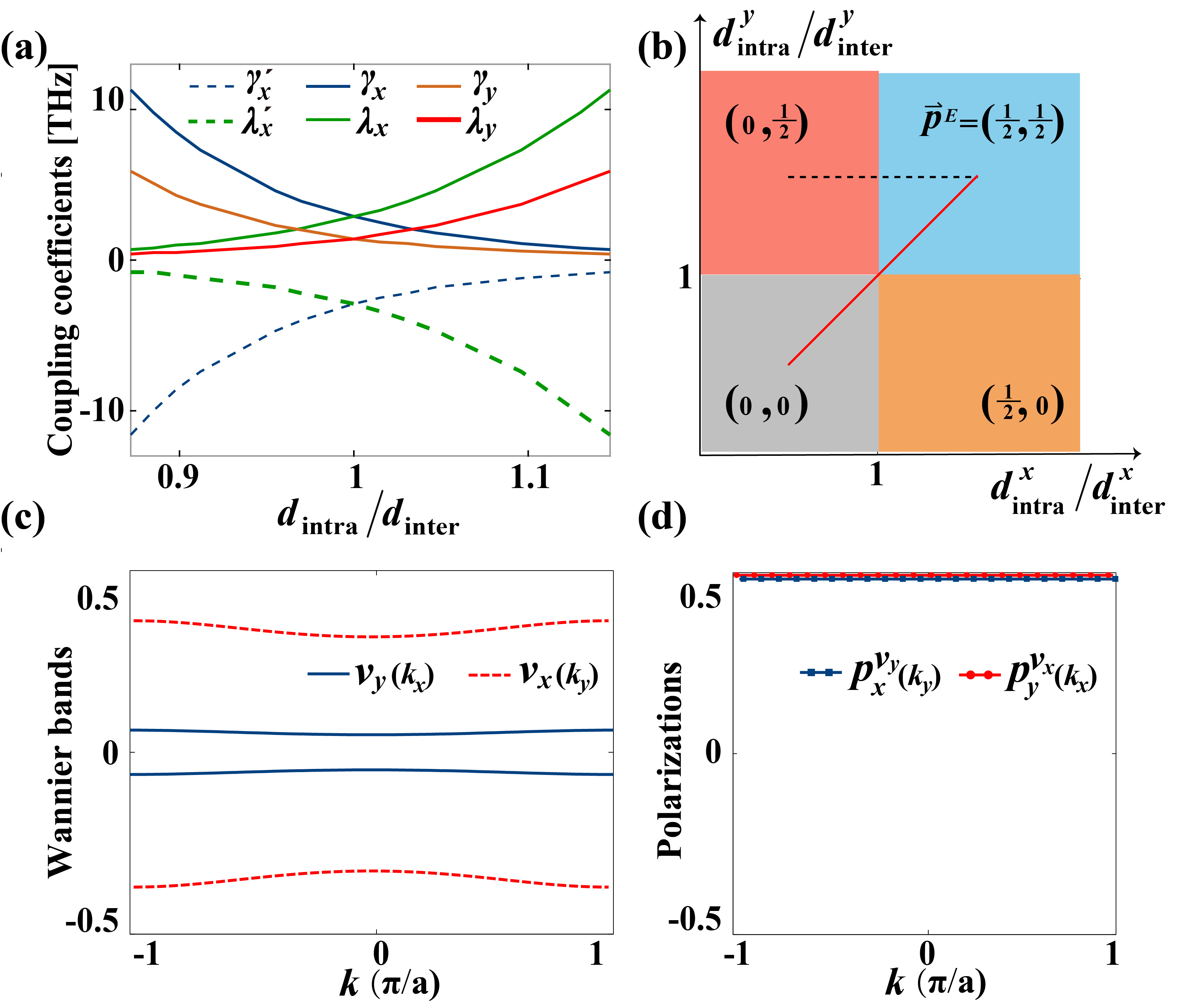}
	\caption{(a) Dependences of the tight-binding coefficients on the ratio $d_\textup{intra}/d_\textup{inter}$ when $d_\textup{intra}^x = d_\textup{intra}^y \equiv  d_\textup{intra}$ with $d_\textup{inter} = a - d_\textup{intra}$. (b) Topological phase diagram for the plasmon-polaritonic system. (c) The Wannier bands and (d) nested Wannier bands for the case with $d_\textup{intra}/d_\textup{inter} = 1.15$ }
	\label{fg2}
\end{figure}
	
In the BBH model, there are four bands where two of them are below the topological band gap. When these two bands are included in the Wilson-loop calculation (i.e., $n,m = 1,2$), the Wilson-loop operator becomes a $2\times2$ matrix. The Wilson-loop operator essentially calibrates the polarization along the y direction. Thus, the eigenvalues of the Wilson-loop operator are the Wannier centers along the y direction, which are obtained by diagonalizing the Wilson-loop operator matrix, $\hat{W}_{y,\bf{k}}{\bf{\eta}}_{y,{\bf{k}}}^j = e^{2\pi{i}v_y^j(k_x)}{\bf{\eta}}_{y,{\bf{k}}}^j$. Here, ${\bf{\eta}}_{y,{\bf{k}}}^j$ is the eigenvector for the $j$-th Wannier band ($j=1,2$). The $j$-th Wannier band is explicitly the dependence of the Wannier center $v_y^j(k_x)$ on the wavevector $k_x$ [see Fig. 2(c)]. The Wannier basis are defined as [1, 2], $|w_j({\bf{k}})\rangle = \sum_{n=1}^2[{\bf{\eta}}_{y,{\bf{k}}}^j]^n|u_n(\bf{k})\rangle$. The nested Wannier band is given by the eigenvalues of the nested Wilson-loop along the $x$ direction for the Wannier functions $p_x^{v_y}(k_y)=\frac{1}{2\pi}\oint{\tilde{A}_1^x({\bf{k}})}$d$k_x$ with $\tilde{A}_1^x({\bf{k}}) = i\langle{w_1({\bf{k}})}|\partial_{k_x}|w_1({\bf{k}})\rangle$ being the Berry connection for the first Wannier band. Similarly, one can calculate the Wannier bands and nested Wannier bands for the Wilson-loop along the $x$ direction and the nested Wilson-loop along the $y$ direction.

The QTIs are featured with two fundamental characteristics: First, the Wannier bands are gapped (meaning that $v_y$ and $v_x$ must not be 0 or 1/2) and symmetric around 0. Such Wannier bands imply that the total bulk dipole polarization vanishes for the topological band, which is a unique feature of QTIs that distinct itself from many other higher-order topological insulators [1, 2]. Second, the Wannier-sector polarizations must be nontrivial, $p_x^{v_y} = p_y^{v_x} = \frac{1}{2}$ where $p_x^{v_y}$ is the average of the nested Wannier bands, i.e., $p_x^{v_y}=\frac{a}{2\pi}\int{dk_y}p_x^{v_y}(k_y)$ and $p_y^{v_x}=\frac{a}{2\pi}\int{dk_x}p_y^{v_x}(k_x)$. It has been demonstrated in Ref. [2] that the Wannier-sector polarizations are equal to the topological edge polarizations induced by the bulk Bloch bands, i.e., $p_x^{E_y} = p_x^{v_y}$ and $p_y^{E_x} = p_y^{v_x}$. As shown in Figs. 2(c) and 2(d), the calculated Wannier bands are indeed gapped and the nested Wannier bands are indeed nontrivially quantized to 1/2 for the QTI phase.

\begin{figure}[h]
	\centering
	\includegraphics[height=5cm]{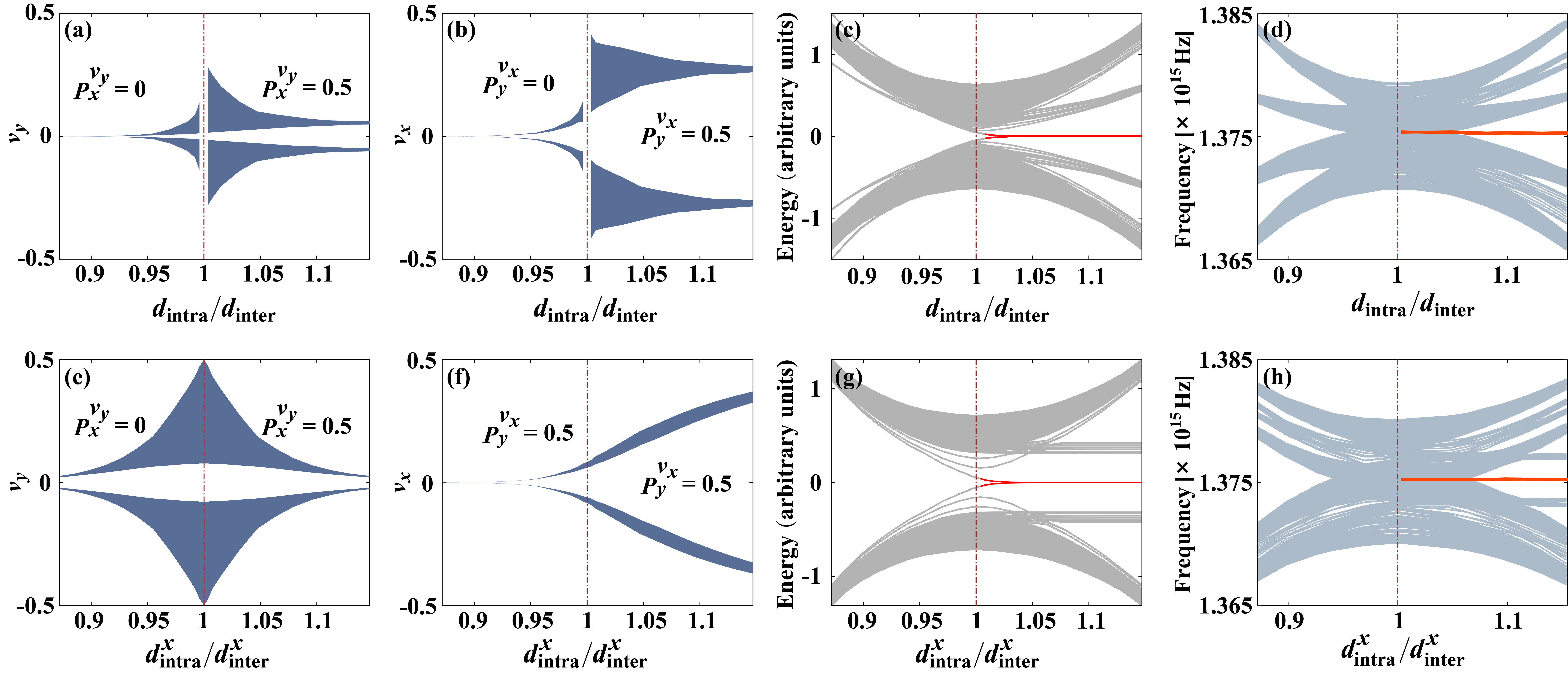}
	\caption{(Color online) (a)-(b) Evolution of the Wannier bands, $v_y$ and $v_x$, along the red line in the phase diagram Fig. 2(b) where $d_\textup{intra}^x = d_\textup{intra}^y \equiv d_\textup{intra} $ and $d_\textup{inter} \equiv a - d_\textup{intra}$. (c)-(d) Evolution of the spectra for a finite-sized supercell with $8\times8$ unit-cells from (c) tight-binding calculation and (d) electrodynamic simulation. (e)-(f) Evolution of the Wannier bands, $v_y$ and $v_x$, along the black line in the phase diagram Fig. 2(b) where $d_\textup{intra}^x$ is changed but  $d_\textup{intra}^y$ is fixed. (g)-(h) Evolution of the spectra in a finite-sized supercell with $8\times8$ unit-cells from (c) tight-binding calculation and (d) electrodynamic simulation. The red lines in those figures indicate the emergence of the corner states in the QTI phase.}
	\label{fg2}
\end{figure}

To reveal the nature of the topological phase transitions in the plasmon-polaritonic systems, we study the evolution of the Wannier bands, Wannier-sector polarizations and the spectrum for a finite-sized (8×8 unit-cells) supercell from both the tight-binding calculation and the electromagnetic-wave simulation. We first study the evolution of those quantities along the red line in the phase diagram Fig. 2(b) where $d_\textup{intra}^x = d_\textup{intra}^y \equiv d_\textup{intra}$ and $d_\textup{inter} \equiv a -d_\textup{intra}$. The topological transition takes place at $d_\textup{intra} = d_\textup{inter} = a/2$ where the bulk band gap closes. We emphasize that for the Wannier-sector polarizations to be well-defined, both the bulk band gap and the Wannier gap are needed. We thus avoid the calculation of the Wannier bands in the vicinity of the phase transition point in Figs. 3(a) and 3(b). In addition, our model is intrinsically anisotropic: even when $d_\textup{intra}^x = d_\textup{intra}^y$, the strength of the tight-binding couplings along the x and y directions are different. Therefore, the Wannier bands, $v_y(k_x)$ and $v_x(k_y)$, are different, as shown in Figs. 3(a) and 3(b) [also in Fig. 2(c)]. Across the topological transition, the Wannier-sector polarizations, $p_x^{v_y}$ and $p_y^{v_x}$,  change from 0 to $\frac{1}{2}$. This is a transition from the trivial phase to the QTI phase, which is consistent with the phase diagram in Fig. 2(b).

The evolution of the spectrum for the supercell with $8\times8$ unit-cells is shown in Fig. 3(c) and 3(d). The former is from the tight-binding calculation, while the latter is from the electromagnetic-wave simulation. The results from these two approaches agree well with each other, confirming the validity of the tight-binding approximation. Here, the spectra include the bulk, edge and corner states. With increasing $d_\textup{intra}/d_\textup{inter}$, the bulk band gap closes and reopen to enter into the QTI phase. The corner states emerge immediately when the topological bulk gap opens, demonstrating directly the higher-order bulk-corner correspondence.

We then study the topological transition along the black line in the phase diagram Fig. 2(b) where $d_\textup{intra}^x$ is changed but with $d_\textup{intra}^y$ fixed. With increasing $d_\textup{intra}^x$, the Wannier band $v_y$ experiences gap closing (i.e., $v_y$ goes to 0.5) and reopening [see Fig. 3(e)], while the Wannier band $v_x$ is kept as gapped [Fig. 3(f)]. Accompanying this transition, the Wannier-sector polarization goes from $p_x^{v_y} = 0$ to $p_x^{v_y} = \frac{1}{2}$, while the  $p_y^{v_x}$ is kept as $\frac{1}{2}$. For the spectrum of the finite-sized supercell, both the tight-binding calculation and the electromagnetic-wave simulation give the same feature that the corner states emerge right after the transition, which confirms the topological phase diagram.

\begin{figure}[h]
	\centering
	\includegraphics[height=7.5cm]{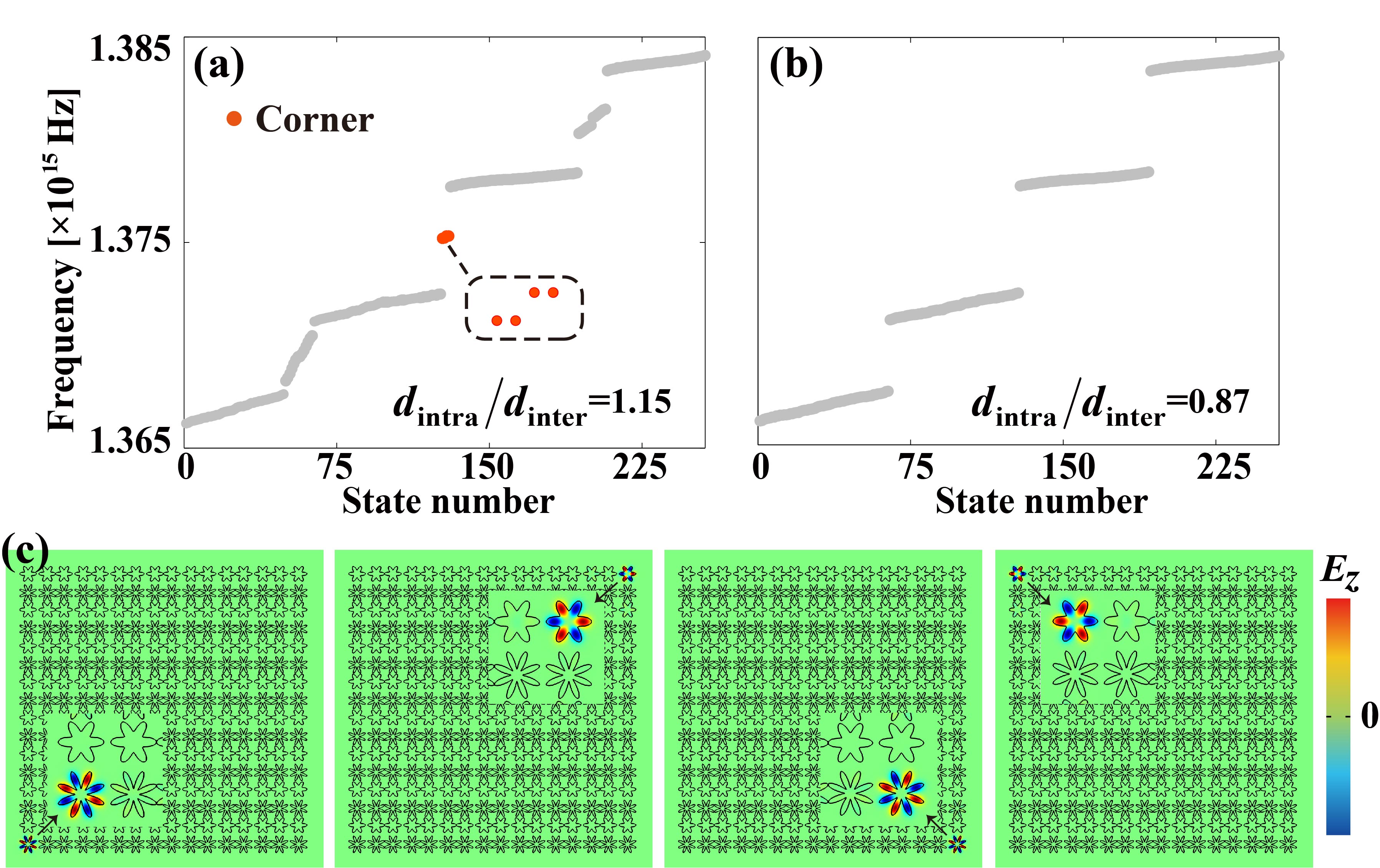}
	\caption{(Color online) (a) Spectrum of the supercell with 8×8 unit-cells for $d_\textup{intra}/d_\textup{inter} = 1.15$ from electromagnetic-wave simulation. The horizontal axis is the number of the eigenmode solution labeled with ascending frequency order. (b) The spectrum for similar supercell but with $d_\textup{intra}/d_\textup{inter} = 0.87$. (c) Distributions of the electric field along the $z$ direction, $E_z$, for the four corner modes in (a). }
	\label{fg4}
\end{figure}

We now show explicitly the emergence of the corner states in the designed plasmon-polaritonic system. Fig. 4(a) presents the spectrum of the supercell with $8\times8$ unit-cells for $d_\textup{intra}/d_\textup{inter} = 1.15$ which is in the QTI phase. There are four corner states emerge in the bulk band gap. In contrast, for the trivial phase with  $d_\textup{intra}/d_\textup{inter} = 0.87$, there is no corner states in the bulk band gap [see Fig. 4(b)]. The electric field profiles of the four corner states for the QTI phase are presented in Fig. 4(c). These results demonstrate convincingly the connection between the quadrupole topology and the in-gap corner states. The edge states for the QTI phase as well as those for the weak topological phases are shown in the Appendix D.

Finally, we remark that because the plasmon-polaritonic system slightly breaks the mirror symmetry along the y direction, i.e., $M_y: y\longrightarrow -y$. The four corner states in Fig. 4(a) are not degenerate, but split into two pairs. In addition, this is also one of the main reasons for the quantitative differences between the tight-binding calculation and the electromagnetic-wave simulation in Fig. 3, although they have the same main features.

\section{Conclusion and outlook.}
We propose a scheme to realize QTIs in plasmon-polaritonic systems. The key ingredient is the sign-reversal mechanism for the couplings between the plasmon-polaritonic cavities from the pole-pole orientation to the node-node orientation. The designed plasmon-polaritonic system demonstrates various topological and trivial phases which can be controlled by the distances between the cavities. Our study introduces quadrupole topology and its topological transitions into plasmonic metamaterials which may inspire future studies on topological metamaterials and their potential applications.

\section*{Acknowledgements}
This work was supported by Jiangsu specially-appointed professor funding and the National Natural Science Foundation of China under the Grant nos. 11675116/11874311; the Fundamental Research Funds for the Central Universities (Grant No. 20720170015). 

\section*{Appendix A: The eigenmodes for six-petaled and eight-petaled cavities}
To obtain the same resonant frequency for hexapole and octopole modes, the parameters of daisy-like cavities ( $r_{01}, d_1, r_{02}$ and $d_2$) and the medium in eight-petaled cavity are finely tuned. In Fig. 5(a), the eigenfrequencies for a single six-petaled cavity and a single eight-petaled cavity are shown, with the parameters in the main text. At the red (hexapole ($f$) mode in six-petaled cavity) and blue (octopole ($g$) mode in eight-petaled cavity) points, the same resonant frequency is obtained at $1.375\times10^{15}$ Hz. In Fig. 5(c), we can see the field patterns of nearby eigenmodes, including the low order modes $p, d$ and high order modes $2s, 2p$. The energy is mainly localized in the nanocavities, with tiny evanescent waves in the background. Fig. 5(b) shows the shifting of octopole mode in the eight-petaled cavity when changing the permittivity $\varepsilon_2$ and keeping other parameters  ( $r_{01}, d_1, r_{02}$ and $d_2$)  unchanged. There is a linear shifting of the eigenfrequency for octopole mode, which intersects with the hexapole mode at $\varepsilon_2=1.525$. The system parameters for the same resonance frequency of six-petaled and eight-petaled cavities are thus achieved. 

\section*{Appendix B: Couplings of eigenmodes in the adjacent cavities}

Here we denote the normalized wavefunctions of $f$ mode in six-petaled cavity as $\psi_1$, while that of $g$ mode in eight-petaled cavity as $\psi_2$, see Fig. 6. Analogizing to the double quantum wells, we put the two cavities at a distance of $a$/2, and the modes of $\frac{1}{\sqrt{2}}(\psi_1-\psi_2)$ and $\frac{1}{\sqrt{2}}(\psi_1+\psi_2)$ are shown in FIG. 6, which have the same features of even-like and odd-like modes in Fig.1(d). Therefore, the coupled eigenmodes in structure III of Fig. 1(d) are exactly the  $\pm\frac{1}{\sqrt{2}}(\psi_1-\psi_2)$ and  $\pm\frac{1}{\sqrt{2}}(\psi_1+\psi_2)$ states, namely, the solutions of double quantum wells.

\section*{Appendix C: Logarithmic fitting between coupling coefficients and $d_\textup{intra}$ ($d_\textup{inter}$)}

In the main text, the changing trends of tight-binding coefficients with ratio $d_\textup{intra}/d_\textup{inter}$ are shown in Fig. 2(a). To further verify the exponential relations, here the logarithm of the tight-binding coefficients are plot to show the trends with $d_\textup{intra}$ and $d_\textup{inter}$. As presented in Fig. 7, the linear relations show the exponential dependence and validate the tight-binding approximation.

The correspondences between $\gamma_x/\lambda_x$ (or $\gamma'_x/\lambda'_x$), $\gamma_y/\lambda_y$ and $d_\textup{intra}/d_\textup{inter}$ are shown in Fig. 8. When $d_\textup{intra}/d_\textup{inter} > 1$, the hopping ratios $\gamma_x/\lambda_x$ ($\gamma_y/\lambda_y$) are less than 1, which corresponds to the topological lattice with $\vec{p}^E = (\frac{1}{2},\frac{1}{2})$.  Conversely, for the region of $d_\textup{intra}/d_\textup{inter} < 1$, the intra-unit-cell coupling $\gamma_x$ ($\gamma_y$) is stronger than the inter-unit-cell coupling $\lambda_x$ ($\lambda_y$), exhibiting trivial characters. Although the blue solid curve has a slight deviation from the other two curves, the overall ratios of them are consistent with each other. Therefore in the main text, we apply the correspondence in Fig. 8 to study the QTI by both the tight-binding method and full-wave simulation. 

\section*{Appendix D: The distributions of edge states under different polarizations }
In Fig. 9, we show the features of the edge states for the QTI phase as well as those for weak topological phases. Firstly, for the QTI phase with $\vec{p}^E = (\frac{1}{2},\frac{1}{2})$, we see that the four corner sites have vanishing field intensity, with the fields locating in the outer boundaries of $x$ or $y$ directions. While for the weak topological phases, i.e., $\vec{p}^E = (0,\frac{1}{2})$, we take $d_\textup{intra}^x = 0.47a$ and $d_\textup{intra}^y = 0.53a$ as an example. As shown in Fig. 8, the edge state along $y$ direction is extended in the whole boundary, and occupies the two corner sites (the edge state on the lower $x$ boundary also has the same features). Moreover, with the increase of $d_\textup{intra}^x$, a phase transition occurs at $d_\textup{intra}^x = 0.5a$ along with the edge band gap closing, as shown in Fig. 3(g). After the transition, the weak topological phases become the QTI phase. Similarly, for the weak topological phases $\vec{p}^E = (\frac{1}{2},0)$, the same edge state emerges along the $x$ direction, as shown in Fig. 9.

\section*{References}

[1] W. A. Benalcazar, B. A. Bernevig and T. L. Hughes. Quantized electric multipole insulators. Science $\bf{357}$, 61 (2017).
\vspace{1ex}  

\noindent{[2] W. A. Benalcazar, B. A. Bernevig and T. L. Hughes. Electric multipole moments, topological multipole moment pumping, and chiral hinge states in crystalline insulators. Phys. Rev. B $\bf{96}$, 245115 (2017).}
\vspace{1ex}  

\noindent{[3] J. Langbehn, Y. Peng, L. Trifunovic, F. von Oppen and P. W. Brouwer. Reflection-Symmetric Second-Order Topological Insulators and Superconductors. Phys. Rev. Lett. $\bf{119}$, 246401 (2017).}
\vspace{1ex}  

\noindent{[4] Z. D. Song, Z. Fang and C. Fang. (d-2)-Dimensional Edge States of Rotation Symmetry Protected Topological States. Phys. Rev. Lett.  $\bf{119}$, 246402 (2017).}
\vspace{1ex}  

\noindent{[5] R. J. Slager, L. Rademaker, J. Zaanen and L. Balents. Impurity-bound states and Green's function zeros as local signatures of topology. Physical Review B  $\bf{92}$, 085126 (2015).} 
\vspace{1ex}  

\noindent{[6] M. Ezawa. Higher-order topological insulators and semimetals on the breathing kagome and pyrochlore SCs. Phys. Rev. Lett.  $\bf{120}$, 026801 (2018).} 
\vspace{1ex}  

\noindent{[7] M. Serra-Garcia, V. Peri, R. Süsstrunk, O. R. Bilal, T. Larsen, L. G. Villanueva and S. D. Huber, Observation of a phononic quadrupole topological insulator. Nature  $\bf{555}$, 342 (2018).} 
\vspace{1ex}  

\noindent{[8] C. W. Peterson, W. A. Benalcazar, T. L. Hughes, and G. Bahl. A quantized microwave quadrupole insulator with topologically protected corner states. Nature  $\bf{555}$, 346 (2018).} 
\vspace{1ex}  

\noindent{[9] S. Imhof, C. Berger, F. Bayer, J. Brehm, L. W. Molenkamp, T. Kiessling, F. Schindler, C. H. Lee, M. Greiter, T. Neupert and R. Thomale. Topolectrical-circuit realization of topological corner modes. Nat. Phys.  $\bf{14}$, 925 (2018).} 
\vspace{1ex}  

\noindent{[10] J. Noh, W. A. Benalcazar, S. Huang, M. J. Collins, K. P. Chen, T. L. Hughes and M. C. Rechtsman. Topological protection of photonic mid-gap defect modes. Nature Photonics  $\bf{12}$, 408 (2018).} 
\vspace{1ex}  

\noindent{[11] F. Schindler, A. M. Cook, M. G. Vergniory, Z. J. Wang, S. S. P. Parkin, B. A. Bernevig and T. Neupert. Higher-order topological insulators. Sci. Adv. $\bf{4}$, eaat0346 (2018).} 
\vspace{1ex}  

\noindent{[12] W. A. Wheeler, L. K. Wagner and T. L. Hughes. Many-body electric multipole operators in extended systems. arXiv:1812.06990 (2018).} 
\vspace{1ex}  

\noindent{[13] B. Y. Xie, H. F. Wang, H. X. Wang, X. Y. Zhu, J. H. Jiang, M. H. Lu and Y. F. Chen. Second-order photonic topological insulator with corner states. Phys. Rev. B. $\bf{98}$, 205147 (2018).}
 \vspace{1ex}  
 
\noindent{[14] H. R. Xue, Y. H. Yang, F. Gao, Y. D. Chong and B. L. Zhang. Acoustic higher-order topological insulator on a kagome lattice. Nat. Mater. $\bf{18}$, 108-112 (2019). }
\vspace{1ex}  

\noindent{[15] X. Ni, M. Weiner, A. Alu and A. B. Khanikaev. Observation of higher-order topological acoustic states protected by generalized chiral symmetry. Nat. Mater. $\bf{18}$, 113-120 (2019). }
\vspace{1ex}  

\noindent{[16] X. Zhang, H. X. Wang, Z. K. Lin, Y. Tian, B. Xie, M. H. Lu, Y.-F. Chen and J.-H. Jiang. Second-order topology and multidimensional topological transitions in sonic crystals. Nat. Phys. $\bf{15}$, 582 (2019).}
\vspace{1ex}  

\noindent{[17] S. Mittal, V. V. Orre, G. Y. Zhu, M. A. Gorlach, A. Poddubny, and M. Hafezi. Photonic quadrupole topological phases. Nat. Photon. $\bf{13}$, 692 (2019).}
\vspace{1ex}  

\noindent{[18] A. E. Hassan, F. K. Kunst, A. Moritz, G. Andler, E. J. Bergholtz, and M. Bourennane. Corner states of light in photonic waveguides. Nat. Photon. $\bf{13}$, 697-700 (2019)}
\vspace{1ex}  

\noindent{[19] H. Y. Fan, B. Z. Xia, L. Tong, S. J. Meng and D. J. Yu. Elastic Higher-Order Topological Insulator with Topologically Protected Corner States. Phys. Rev. Lett. $\bf{122}$, 204301 (2019).}
\vspace{1ex}  

\noindent{[20] Y. Ota, F. Liu, R. Katsumi, K. Watanabe, K. Wakabayashi, Y. Arakawa and S. Iwamoto. Photonic crystal nanocavity based on a topological corner state. Optica $\bf{6}$, 786 (2019). }
\vspace{1ex}  

\noindent{[21] X. D. Chen, W. M. Deng, F. L. Shi, F. L. Zhao, M. Chen and J. W. Dong. Direct Observation of Corner States in Second-Order Topological Photonic Crystal Slabs. Phys. Rev. Lett. $\bf{122}$, 233902 (2019). }
\vspace{1ex}  

\noindent{[22] B. Y. Xie, G. X. Su, H. F. Wang, H. Su, X. P. Shen, P. Zhan, M. H. Lu, Z. L. Wang and Y. F. Chen. Visualization of Higher-Order Topological Insulating Phases in Two-Dimensional Dielectric Photonic Crystals. Phys. Rev. Lett. $\bf{122}$, 233903 (2019). }
\vspace{1ex}  

\noindent{[23] L. Zhang, Y. Yang, P. Qin, Q. Chen, F. Gao, E. Li, J.-H. Jiang, B. Zhang, and H. Chen. Higher-order photonic topological states in surface-wave photonic crystals. arXiv:1901.07154 (2019).}
\vspace{1ex}  

\noindent{[24] Y. Chen, X. C. Lu and H. Y. Chen. Effect of truncation on photonic corner states in a Kagome lattice. Opt. Lett. $\bf{44}$, 4251 (2019).}
\vspace{1ex}  

\noindent{[25] Z. K. Lin, H. X. Wang, M. H. Lu and J. H. Jiang. Nonsymmorphic topological quadrupole insulator in Sonic Crystals. arXiv:1903.05997 (2019).}
\vspace{1ex}  

\noindent{[26] B. Kang, K. Shiozaki, and G. Y. Cho. Many-Body Order Parameters for Multipoles in Solids. arXiv:1812.06999}
\vspace{1ex}  

\noindent{[27] M. Z. Hasan and C. L. Kane. Colloquium: topological insulators. Rev. Mod. Phys. $\bf{82}$, 3045 (2010).}
\vspace{1ex}  

\noindent{[28] X. L. Qi and S. C. Zhang. Topological insulators and superconductors. Rev. Mod. Phys. $\bf{83}$, 1057 (2011).}
\vspace{1ex}  

\noindent{[29] Z. Wang, Y. Chong, J. D. Joannopoulos, and M. Soljacic. Observation of unidirectional backscattering immune topological electromagnetic states. Nature $\bf{461}$, 772-775 (2009).}
\vspace{1ex}  

\noindent{[30] M. Hafezi, E. A. Demler, M. D. Lukin, and J. M. Taylor. Robust optical delay lines with topological protection. Nat. Phys. $\bf{7}$, 907-912 (2011).}
\vspace{1ex}  

\noindent{[31] Y. Poo, R. Wu, Z. Lin, Y. Yang, and C. T. Chan. Experimental realization of self-guiding unidirectional electromagnetic edge states. Phys. Rev. Lett. $\bf{106}$, 093903 (2011).}
\vspace{0.5ex} 
 
\noindent{[32] Y. E. Kraus, Y. Lahini, Z. Ringel, M. Verbin, and O. Zilberberg. Topological states and adiabatic pumping in quasicrystals. Phys. Rev. Lett. $\bf{109}$, 106402 (2012).}
\vspace{1ex}  

\noindent{[33] M. Hafezi, S. Mittal, J. Fan, A. Migdall, and J. M. Taylor. Imaging topological edge states in silicon photonics. Nat. Photon. $\bf{7}$, 1001-1005 (2013).}
\vspace{1ex}  

\noindent{[34] A. B. Khanikaev, et al. Photonic topological insulators. Nat. Mater. $\bf{12}$, 233–239 (2013).}
\vspace{1ex}  

\noindent{[35] M. C. Rechtsman, et al. Photonic Floquet topological insulators. Nature $\bf{496}$, 196-200 (2013). }
\vspace{1ex}  

\noindent{[36] W.-J. Chen, et al. Experimental realization of photonic topological insulator in a uniaxial metacrystal waveguide. Nat. Commun. $\bf{5}$, 6782 (2014). }
\vspace{1ex}  

\noindent{[37] L. Lu, J. D. Joannopoulos, and M. Soljacic. Topological photonics. Nat. Photon. $\bf{8}$, 821-829 (2014).}
\vspace{1ex}  

\noindent{[38] X. Cheng, et al. Robust reconfigurable electromagnetic pathways within a photonic topological insulator. Nat. Mater. $\bf{15}$, 542-548 (2016).}
\vspace{1ex}  

\noindent{[39] L. H. Wu and X. Hu. Scheme for Achieving a Topological Photonic Crystal by Using Dielectric Material. Phys. Rev. Lett $\bf{114}$, 223901 (2015).}
\vspace{1ex}  

\noindent{[40] L. Xu, H. X. Wang, Y. D. Xu, H. Y. Chen and J. H. Jiang. Accidental degeneracy in photonic bands and topological phase transitions in two-dimensional core-shell dielectric photonic crystals. Opt. Express $\bf{24}$, 18059 (2016).}
\vspace{1ex}  

\noindent{[41] A. B. Khanikaev and G. Shvets. Two-dimensional topological photonics. Nat. Photon. $\bf{11}$, 763 (2017).}
\vspace{1ex}  

\noindent{[42] H.-X. Wang, Y. Chen, Z. H. Hang, H.-Y. Kee, and J.-H. Jiang. Type-II Dirac photons.
npj Quantum Mater. $\bf{2}$, 54 (2017).}
\vspace{1ex}  

\noindent{[43] X. Zhu, H.-X. Wang, C. Xu, Y. Lai, J.-H. Jiang, and S. John. Topological transitions in continuously deformed photonic crystals. Phys. Rev. B $\bf{97}$, 085148}
\vspace{1ex}  

\noindent{[44] Y. Chen, L. Xu, G. X. Cai and H. Y. Chen. Chemical bonds and edge states in a metamolecular crystal. Phys. Rev. B $\bf{98}$, 125430(2018).}
\vspace{1ex}  

\noindent{[45] T. Ozawa, H. M. Price, A. Amo, N. Goldman, M. Hafezi, L. Lu, M. C. Rechtsman, D. Schuster, J. Simon, O. Zilberberg and I. Carusotto. Topological photonics. Rev. Mod. Phys. $\bf{91}$, 015006 (2019).}
\vspace{1ex}  

\noindent{[46] S. A. Maier. Plasmonics: fundamentals and applications (Springer, 2007).}
\vspace{1ex}  

\noindent{[47] A. Boltasseva and H. A. Atwater. Low-Loss Plasmonic Metamaterials. Science $\bf{331}$, 290 (2011).}
\vspace{1ex}  

\noindent{[48] K. Yao and Y. M. Liu. Plasmonic metamaterials. Nanotechnology Reviews $\bf{3}$, 177 (2014). }
\vspace{1ex}  

\noindent{[49] F. Gao, Z. Gao, X. H. Shi, Z. J. Yang, X. Lin, H. Y. Xu, J. D. Joannopoulos, M. Soljacic, H. S. Chen, L. Lu, Y. D. Chong and B. L. Zhang. Probing topological protection using a designer surface plasmon structure. Nat. Commun. $\bf{7}$, 11619 (2016). }
\vspace{1ex}  

\noindent{[50] S. Yves, R. Fleury, T. Berthelot, M. Fink, F. Lemoult and G. Lerosey. Crystalline metamaterials for topological properties at subwavelength scales. Nat. Commun. $\bf{8}$, 16023 (2017).}
\vspace{1ex}  

\noindent{[51] X. X. Wu, Y. Meng, J. X. Tian, Y. Z. Huang, H. Xiang, D. Z. Han and W. J. Wen. Direct observation of valley-polarized topological edge states in designer surface plasmon crystals. Nat. Commun. $\bf{8}$, 1304 (2017). }
\vspace{1ex}  

\noindent{[52] Z. Gao, L. Wu, F. Gao, Y. Luo and B. L. Zhang. Spoof plasmonics: from metamaterial concept to topological description. Adv. Mater. $\bf{30}$, 1706683 (2018).}
\vspace{1ex}  

\noindent{[53] Y. Chen, H. Y. Chen, and G. X. Cai. High transmission in a metal-based photonic crystal. Appl. Phys. Lett. $\bf{112}$, 013504 (2018).}
\vspace{1ex}  

\noindent{[54] T. Liu, Y. R. Zhang, Q. Ai, Z. P. Gong, K. Kawabata, M. Ueda and F. Nori. Second-Order Topological Phases in Non-Hermitian Systems. Phys. Rev. Lett. $\bf{122}$, 076801 (2019).}
\vspace{1ex}  

\noindent{[55] Z. Gao, F. Gao, Y. M. Zhang, H. Y. Xu, Y. Luo and B. L. Zhang. Forward/Backward Switching of Plasmonic Wave Propagation Using Sign-Reversal Coupling. Adv. Mater. $\bf{29}$, 1700018 (2017).}

\begin{figure}[h]
	\centering
	\includegraphics[height=8cm]{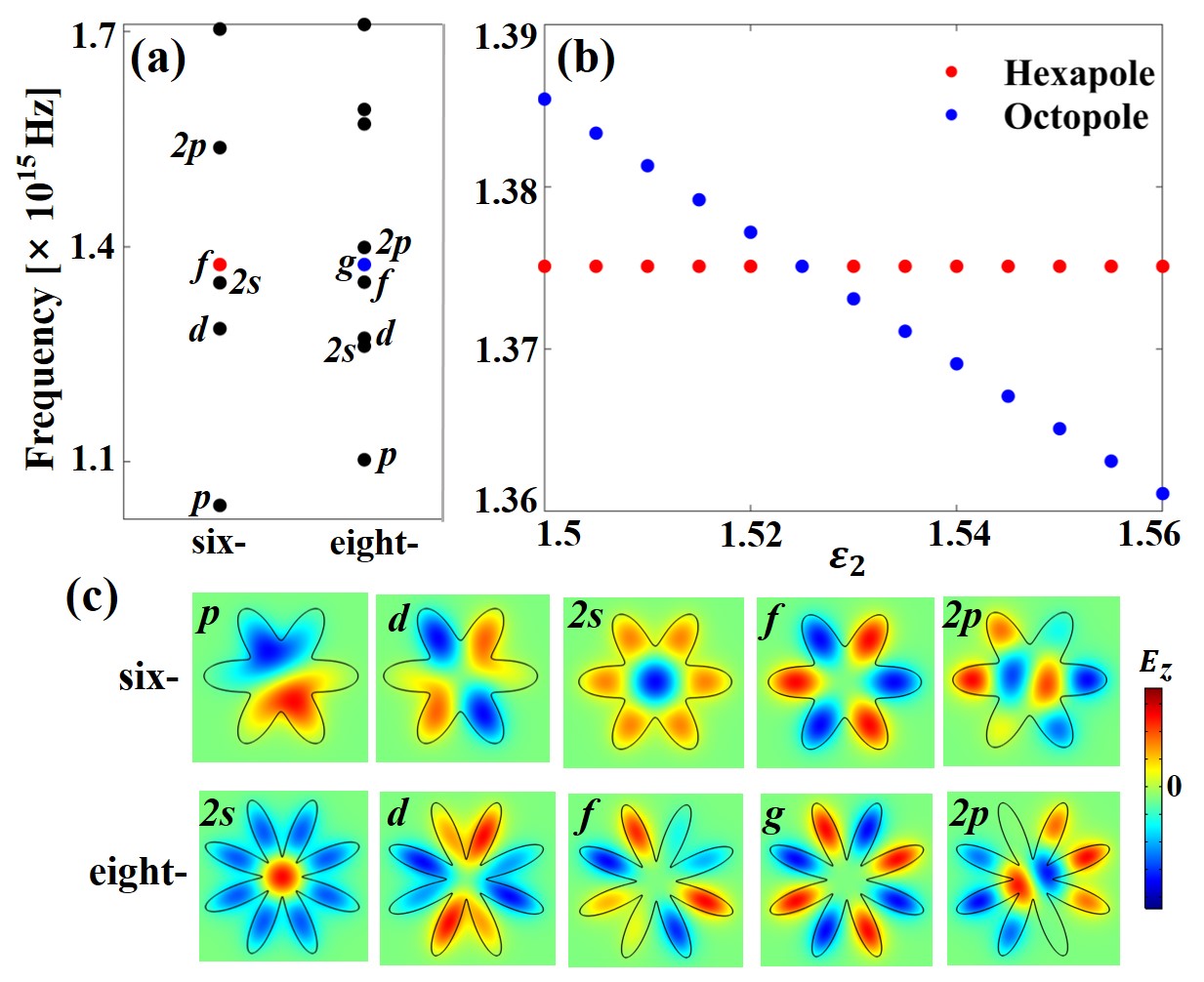}
	\caption{(a) Partial eigenfrequencies for six-petaled and eight-petaled cavities at the parameters in the main text, where hexapole (red point) and octopole modes (blue point) have the same frequency. (b) Changing of the eigenfrequency at octopole mode when tuning the permittivity in the eight-petaled cavity. (c) The field patterns $E_z$ for partial eigenmodes in (a). }
	\label{fg5}
\end{figure}

\begin{figure}[h]
	\centering
	\includegraphics[height=6cm]{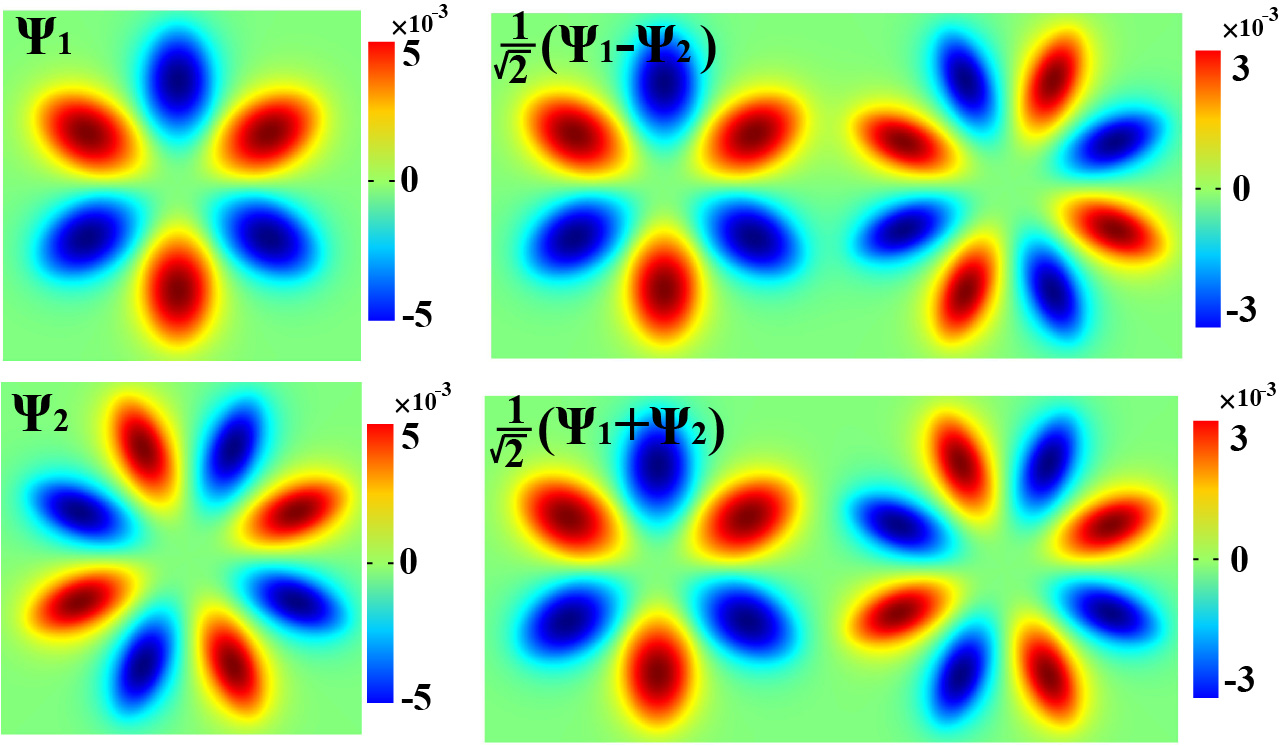}
	\caption{Normalized wavefunctions of $f$ mode $(\psi_1)$ in a single six-petaled and $g$ mode $(\psi_2)$ in a single eight-petaled cavity. The eigenmodes of $\frac{1}{\sqrt{2}}(\psi_1-\psi_2)$ and $\frac{1}{\sqrt{2}}(\psi_1+\psi_2)$ when the two cavities are placed at a distance of $a$/2.   }
	\label{fg5}
\end{figure}

\begin{figure}[h]
	\centering
	\includegraphics[height=6.5cm]{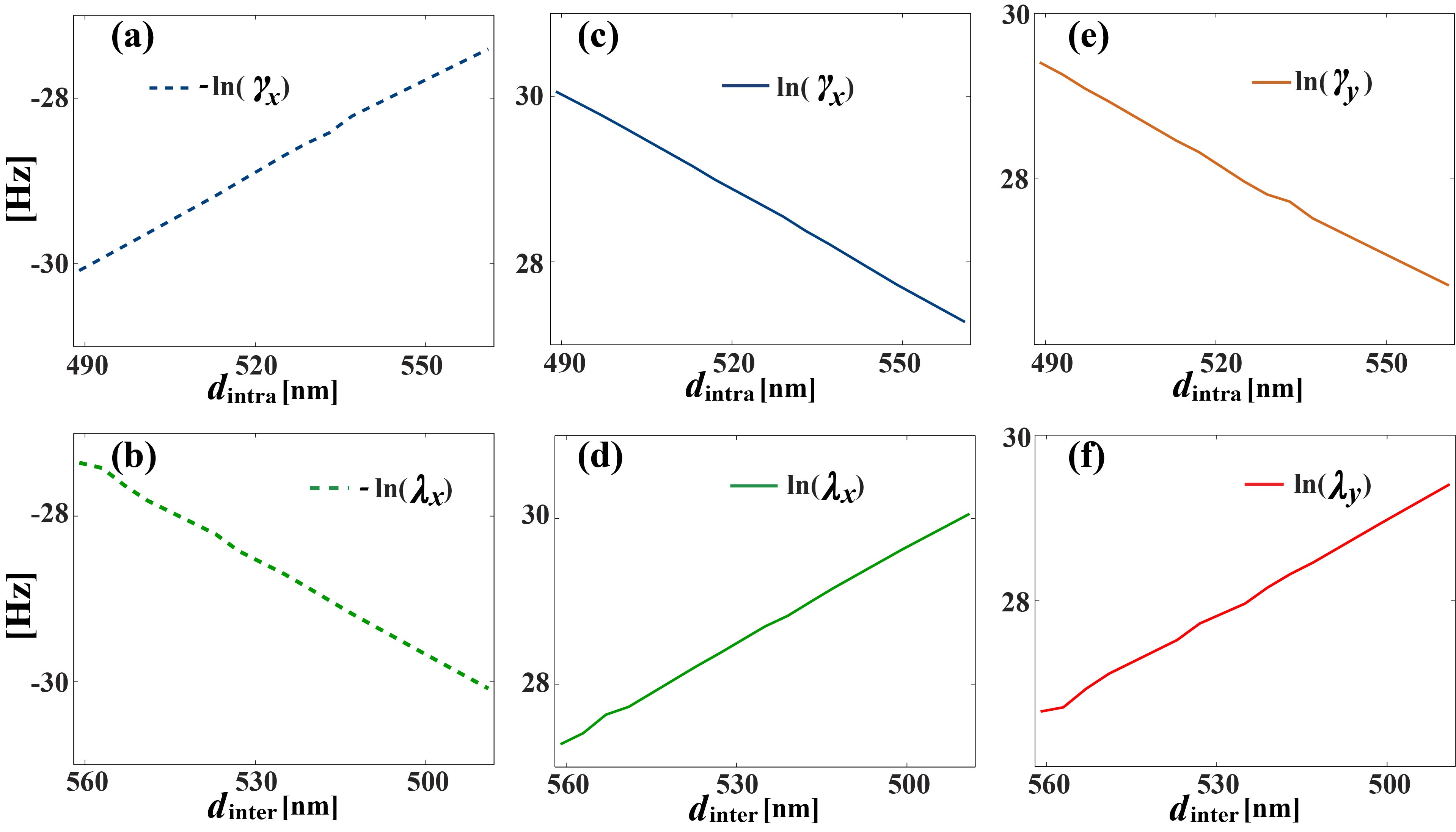}
	\caption{ Relations between the logarithm of coupling coefficients and the distances $d_\textup{intra}$ and $d_\textup{inter}$ in the unit cell, where all the six curves in FIG. 2(a) are plotted.}
	\label{fg6}
\end{figure}

\begin{figure}[h]
	\centering
	\includegraphics[height=7cm]{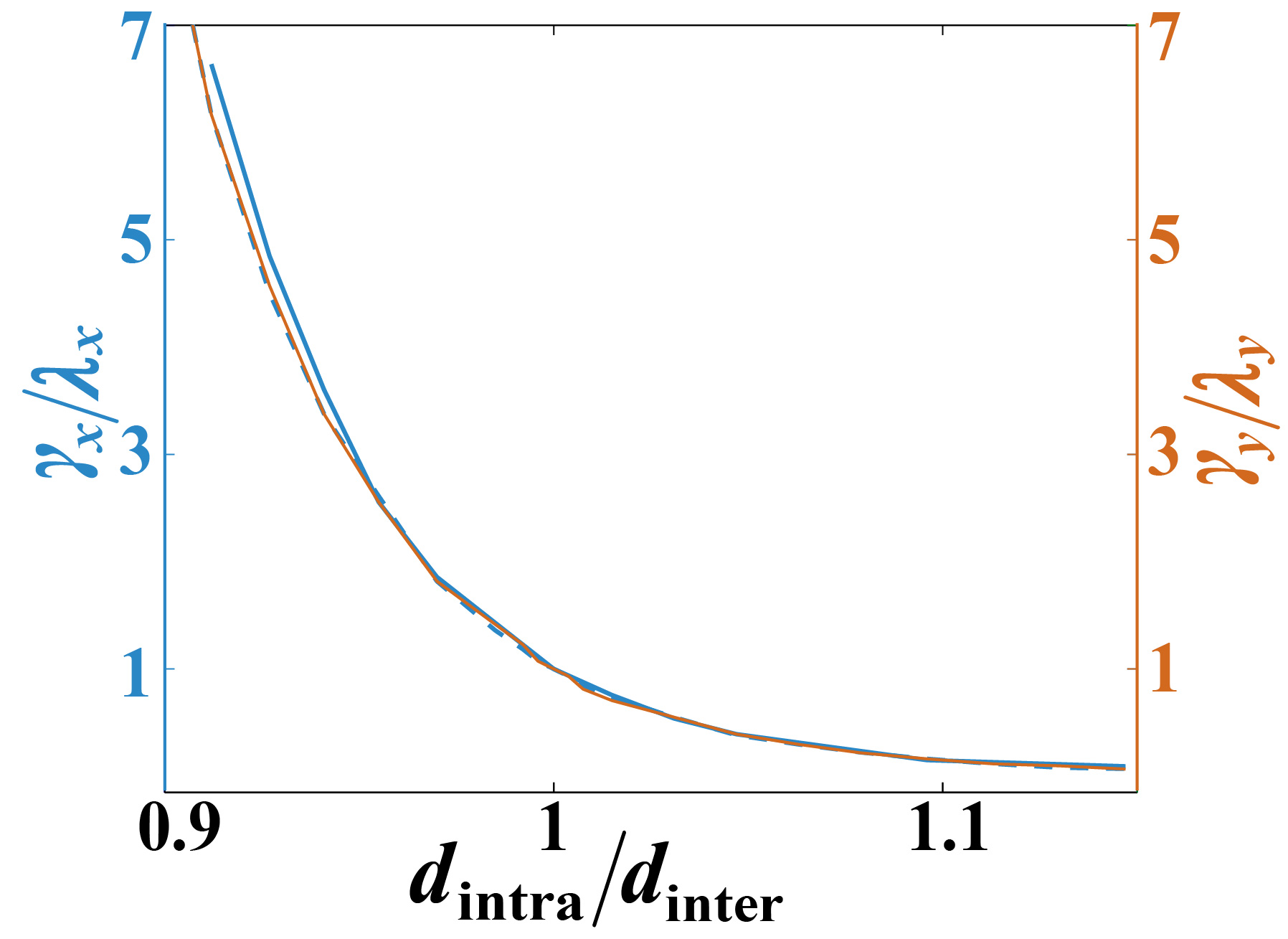}
	\caption{ Correspondence between the ratios of $d_\textup{intra}/d_\textup{inter}$ in this plasmon-polaritonic system and $\gamma_x/\lambda_x$, $\gamma_y/\lambda_y$ in tight-binding model, when  $d_\textup{intra}^x = d_\textup{intra}^y \equiv d_\textup{intra} $. }
	\label{fg7}
\end{figure}

\begin{figure}[h]
	\centering
	\includegraphics[height=8cm]{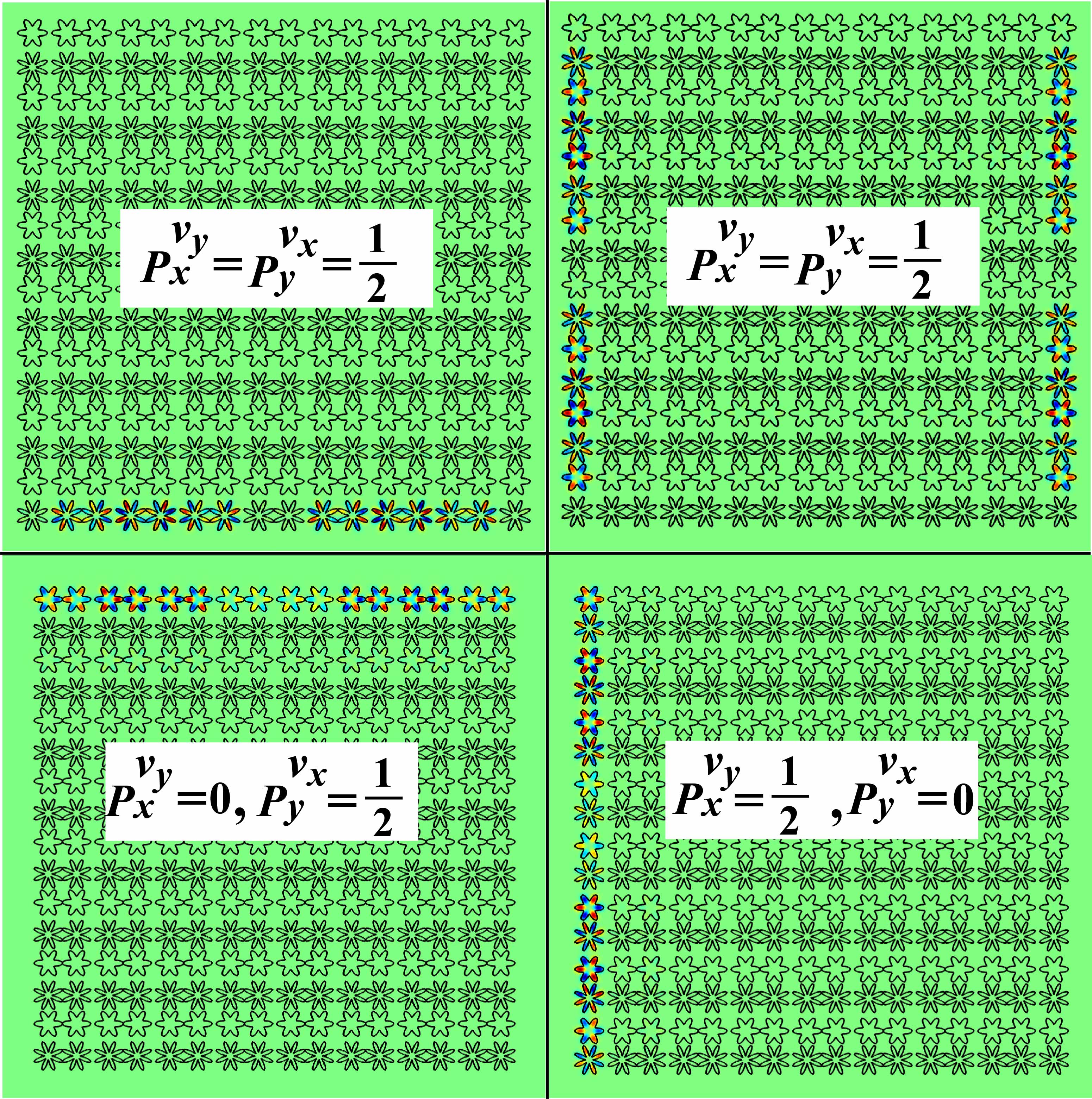}
	\caption{The distributions of edge states for the QTI phase as well as the weak topological phases.}
	\label{fg7}
\end{figure}

\end{document}